# Smoking effect on the circadian rhythm of blood pressure in hypertensive subjects

G. Silveri[1], L. Pascazio[2], A. Miladinović[1], M. Ajčević[1] and A. Accardo[1]
[1] Department of Engineering and Architecture, University of Trieste, Trieste 34127, Italy
[2] Department of Medical, Surgical and Health Care, CS of Geriatrics, University of Trieste, Trieste 34127, Italy

*Abstract*—The use of office measurement of Blood Pressure (BP) as well as of the mean on day-time, on night-time or on 24h does not accurately describe the changes of the BP circadian rhythm. Moreover, several risk factors affect this rhythm but until now possible alterations, due to the presence of such risk factors considered separately, were not been yet studied. Cigarette smoking is one of the most relevant risk factors increasing cardiovascular morbidity and mortality. The aim of this study is to evaluate quantitatively and with a suitable temporal detail how the smoking influences the BP circadian rhythm in normotensive and hypertensive subjects excluding those who presented other risk factors like obesity, dyslipidemia and diabetes mellitus. Holter BP monitoring coming from 618 subjects was used and the behaviour on 24h was examined separately in normotensive and hypertensive subjects either smokers or non-smokers. Four intervals with alternate different characteristics were found in the BP rhythm and regression lines approximated them in order to evaluate the changing rate of BP in each period. Results showed higher values from 10:00 to 02:00 in hypertensive smokers than non-smokers and significant differences between normotensive smokers and non-smokers between 10:00 and 19:00. The changing rate between 10:00 and 14:30 was higher in non-smokers than in smokers for both normotensive and hypertensive subjects while the opposite was found in the other three periods. The different velocity rates of BP changes during 24h, could be associated with different risk levels of cardiovascular disease.

## I. INTRODUCTION

BLOOD pressure (BP) is a biological signal that has a circadian rhythm related to the cardiovascular system. The development of cardiovascular disease is mainly associated with the increase of some risk factors like high-density lipoprotein-cholesterol ratio, excess weight, elevated blood sugar level and hypertension. The latter is frequently associated with the incidence of stroke, myocardial infarction, heart failure and death. In hypertensive population, smoking increases the development of cardiovascular events [1]-[2] due to endothelial dysfunction, producing organ damages [3]-[4]. Moreover, it is known that to smoke a cigarette produces, because of vasoconstriction, an acute increase in blood pressure in normotensive subjects for at least 15 minutes, with a peak elevation ranging from 3 to 12 mmHg of systolic blood pressure (SBP) and from 5 to 10 mmHg of diastolic one (DBP) [5]. Moreover, in an epidemiological study, Al-Safi found that normotensive smokers had significant higher office blood pressure compared with non-smokers [6]. However, in order to better evaluate the subjects' physiological condition and to correctly diagnose and manage hypertension, the 24h Ambulatory BP Monitoring (ABPM) has proven to be more appropriate than the conventional office BP measure [7]. Thus, other studies related to 24h BP monitoring found that, even though office BP levels were comparable between normotensive smokers and non-smokers, ABPM in smokers maintained a higher day-time systolic value than in non-smokers [8]-[9]. This is in contrast with the results of Mikkelsen et al. [10], in which they highlighted in a cohort of normotensive adults from 20 to 79 years old, that smokers exhibited slightly lower BP ambulatory blood pressure compared with non-smokers during day and night.

Furthermore, also in hypertensive subjects smoking has been shown to induce an acute rise in blood pressure in office condition [11]. Comparing hypertensive smokers and non-smokers, Bang et al. found that there was no difference between smokers and non-smokers in office conditions. On the contrary, by using ABPM, they underlined that smokers had a significant higher systolic and diastolic blood pressure in the period between 06:00 and 22:00 (day-time). In addition, they highlighted that the difference between office and day-time ABPM was significant and lower only for diastolic BP in the hypertensive smokers compared to non-smokers [12]. Some studies [9], [13] confirmed a higher day-time BP level in hypertensive smokers compared to hypertensive non-smokers; other authors [14] highlighted that also the mean on 24h was significantly higher in hypertensive smokers while the difference between smokers and non-smokers in the night-time (22:00 to 06:00) was not significant.

However, the use of a single value (office as well as mean on day-time, on night-time or on 24h) only approximatively describes the changes due to the circadian rhythm of BP occurring during 24h. A better temporal definition of the changes occurring in BP during 24h can also be useful to identify with greater accuracy not only possible pressure peaks but also how quickly changes occur in BP.

Therefore, in this study, we examine with a fine temporal resolution how the BP changes along 24h in normotensive and hypertensive smokers compared to non-smokers, in intervals of 15 or 30 minutes. Moreover, to reduce the effects of other cardiovascular risk factors such as obesity, dyslipidemia and diabetic mellitus, which generally are positively associated with BP, increasing its value [15]-[16], we studied hypertensive and normotensive subjects did not present these risk factors.

## II. MATERIALS AND METHODS

The study population consisted of 618 subjects afferent to the Geriatric Department of the Trieste, from June 2016 to September 2016. This retrospective study was carried out on subjects who met these inclusion criteria: no clinical or laboratory evidence of secondary arterial hypertension, absence of clinical evidence of hypertension-related complications and no cardiac disease. According to current guidelines, subjects were classified based on office BP readings, as hypertensive (SBP≥140 mmHg and/or DBP≥90 mmHg) or normotensive (SBP<140 mmHg and DBP<90 mmHg) [17]. The smoking, obesity, diabetes mellitus and dyslipidemia risk factors were collected on physical examination at the time of visit, in accordance with international guidelines [18]. We excluded subjects that presented at least one of these risk factors; the remaining subjects (Table I) were divided into 32 hypertensive smokers (HS) and 113 non-smokers (HNS), 20 normotensive smokers (NHS) and 83 non-smokers (NHNS) presenting comparable age. Non-smokers were defined as those who had not smoked for at least a year. The measurement procedures were in accordance with the institutional guidelines and the principles of Helsinki Declaration. All the subjects gave their informed consent.

TABLE I
SUBJECT GROUPS

| Groups | Age | Gender | |
|---|---|---|---|
| | | Male | Female |
| NHNS | 59±16 | 46 | 37 |
| NHS | 54±15 | 13 | 7 |
| HNS | 56±13 | 58 | 55 |
| HS | 52±14 | 18 | 14 |

The BP was measured in office condition around at 9:00, as the average of two consecutive readings and then in ambulatory way, along the 24h, by using a Holter Blood Pressure Monitor (Mobil-O-Graph® NG, IEM gmbh Stolberg, Germany) [17]. The portable monitor was programmed to obtain ambulatory blood pressure each 15-min interval throughout the day-time (6:00 to 22:00) and each 30-min interval throughout the night-time (22:00 to 6:00). No patient received additional medication that might affect the circadian blood pressure. Written instruction was given on how to use the Holter Blood Pressure Monitor was given to participants individually.

The circadian trend of the mean values of SBP and DBP was separately examined for H, HS, NHS and NHNS subject groups. Since the BP profiles during 24h showed to be bimodal with two maxima and two minima, we subdivided the 24h into four intervals. In each period the quite linear trend was fitted by a regression line. Slopes, intercepts, $R^2$ and p-value parameters were calculated for each line. Finally, we compared the trends between normotensive smokers and non-smokers and between hypertensive smokers and non-smokers for each interval, by using the Wilcoxon rank sum test (p<0.05).

III. RESULTS

The median values (±1SD) of BP in the four subject groups for each period as well as during day- and night-time are reported in Table II. The differences between circadian SBP and DBP values in hypertensive smokers compared with non-smokers were significant in the three periods between 10:00 and 02:00 (p<0.05) and also considering the day-time from 06:00 to 22:00 (p<0.0001). Moreover, in normotensive subjects, the difference between smokers and non-smokers was significant only for SBP values between 10:00 and 19:00 (p<0.04), as well as in day-time and night-time (p<0.003).

Figure 1 shows the circadian rhythms of SBP and DBP among the subjects in the four groups and the linear approximations in the four periods in which we separated each rhythm. All the trends decrease between 10:00 and 14:30 and between 19:15 and 02:00, while the increase from 14:45 to 19:00 and from 2:30 to 09:45 and their behavior was approximated, both for SBP and DBP, by linear regressions (Fig.1) presenting slopes that differ among groups and intervals (Tab III).

TABLE II
25TH AND 75TH PERCENTILES OF BP PRESSURES (IN MMHG) IN THE FOUR SUBJECT GROUPS FOR EACH PERIOD AND DURING DAY- AND NIGHT-TIMES. DIFFERENCES BETWEEN SMOKERS AND NON-SMOKERS: #P<0.04, *P<0.05, §P<0.003, $P<0.0001

| Systolic BP | NHNS | NHS | HNS | HS |
|---|---|---|---|---|
| 10:00-14:30 | 124-128# | 121-126# | 137-143* | 142-146* |
| 14:45-19:00 | 124-126# | 121-125# | 134-137* | 136-144* |
| 19:15-02:00 | 116-127 | 112-125 | 121-138* | 126-144* |
| 02:30-09:45 | 113-122 | 109-122 | 120-139 | 120-134 |
| Day-time | 122-127§ | 120-125§ | 134-140$ | 136-145$ |
| Night-time | 113-116§ | 107-113§ | 118-121 | 118-125 |
| Dyastolic BP | | | | |
| 10:00-14:30 | 77-81 | 77-81 | 88-90* | 88-93* |
| 14:45-19:00 | 76-78 | 77-81 | 83-86* | 85-91* |
| 19:15-02:00 | 68-79 | 71-79 | 72-85* | 75-91* |
| 02:30-09:45 | 66-75 | 68-79 | 73-86 | 72-84 |
| Day-time | 75-80 | 77-81 | 83-88$ | 85-92$ |
| Night-time | 66-68 | 67-71 | 69-74 | 70-77 |

In particular, the slopes were generally greater for SBP in smokers than in non-smokers (in both hypertensive and normotensive subjects) in the three periods after 14.45 while an opposite trend was present between 10.00 and 14.30 with slope values greater in non-smokers. For DBP the slopes were quite similar among the subject groups, excluded HS, in the three periods after 14.45 while from 14.45 to 02.00 in the HS group, the slopes were much higher. As for SBP, also for DBP, in the first period, between 10.00 and 14.45, the slopes were greater in non-smokers than in smokers.

TABLE III
SLOPES ($m$ IN MMHG/HOUR) AND INTERCEPTS ($q$ IN MMHG) OF THE REGRESSION LINES BETWEEN HOURS OF THE DAY AND BP, IN THE FOUR GROUPS FOR EACH PERIOD. *NO SIGNIFICANTLY DIFFERENT FROM ZERO.

| | NHNS | | NHS | | HNS | | HS | |
|---|---|---|---|---|---|---|---|---|
| Systolic BP | m | q | m | q | m | q | m | q |
| 10:00-14:30 | -2.01 | 151 | -0.52* | 130 | -3.13 | 179 | -2.32 | 173 |
| 14:45-19:00 | 1.03 | 108 | 2.04 | 88 | 1.06 | 117 | 2.92 | 90 |
| 19:15-02:00 | -2.62 | 180 | -3.71 | 201 | -3.77 | 213 | -4.25 | 229 |
| 02:30-09:45 | 2.12 | 53 | 2.77 | 31 | 4.32 | -4 | 3.71 | 14 |
| Dyastolic BP | m | q | m | q | m | q | m | q |
| 10:00-14:30 | -1.67 | 99 | -0.58* | 86 | -2.37 | 118 | -2.15 | 118 |
| 14:45-19:00 | 1.13 | 59 | 1.69 | 50 | 1.13 | 65 | 2.95 | 38 |
| 19:15-02:00 | -2.52 | 129 | -2.34 | 127 | -3.28 | 152 | -4.04 | 173 |
| 02:30-09:45 | 2.15 | 6 | 2.24 | 6 | 3.39 | -24 | 3.22 | -20 |

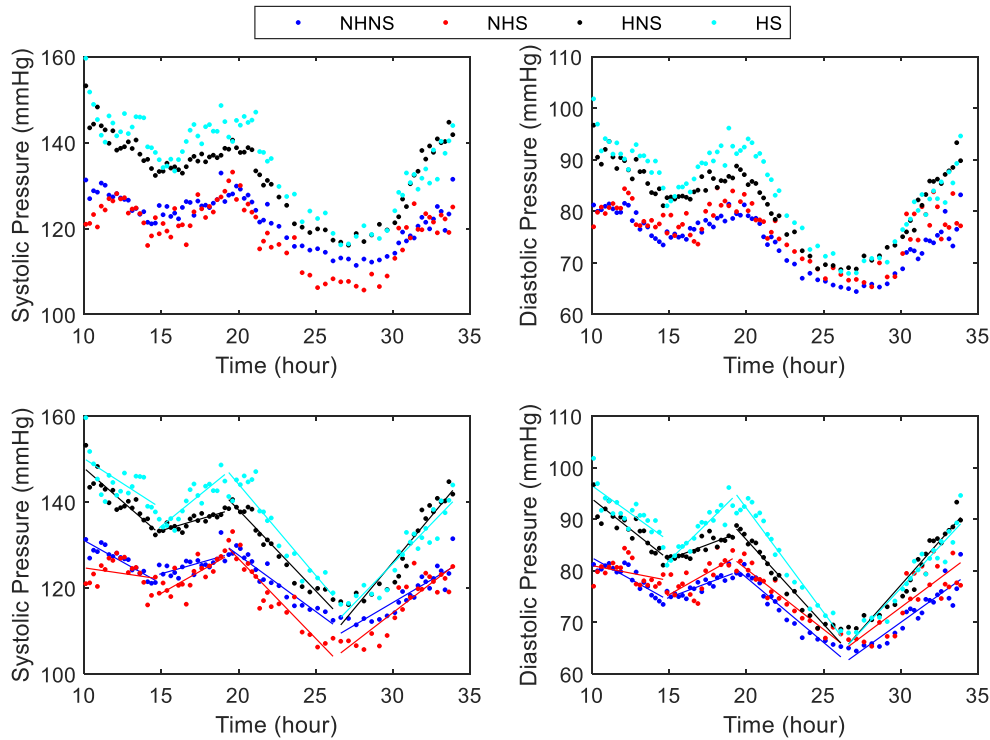

Fig. 1: Circadian Systolic and Diastolic BP rhythms in Hypertensive Smokers (HS), Hypertensive Non-Smokers (HNS), Normotensive Smokers (NHS) and Normotensive Non-Smokers (NHNS) together with the linear approximations in the four separate periods.

R-squared determination coefficient presented values between 0.40 and 0.95 for both SBP and DBP, confirming a good linear approximation. The only interval in which the BP presented a quite constant behaviour was between 10:00 and 14:30 in normotensive smokers. In all the other cases the slopes were significantly ($p<0.01$) different from zero.

## IV. DISCUSSION

In this study, we considered subjects eventually affected by one cardiovascular risk factor (smoking) not taking into account other risk factors that could modify the circadian BP rhythms. Although the effects of smoking on BP are complex with evidence that it increases the risk of renovascular, malignant and masked hypertension [19], to our knowledge no other authors have studied the influence of smoking on the circadian rhythm separating the effects of other risk factors. Thus, our study is the first that evaluate the effect of smoking on the circadian BP rhythm. To assess BP values, we used the ABPM that is more predictive than office BP in estimating cardiovascular risks because of a more through information.

The study pointed out four well definite periods during 24h in hypertensive as well as in normotensive smokers and non-smokers. One of the main findings was that the normotensive subjects presented higher mean values of SBP in non-smokers than in smokers confirming the result of Mikkelsen [10] although opposite to those of [8]-[9], during day-time. The lower BP values in normotensive smokers presenting between 10:00 to 19:00 cannot be justified by the known pharmacological effect of nicotine that increases the BP [20]. One possible explanation for the lower BP in normotensive smokers could be the tolerance developed of the nicotine effects [18]. On the other hand, hypertensive smokers presented significant higher values than non-smokers along the periods between 10:00 and 02:00 while during the awaking, the behaviour was opposite. These results confirm the findings of numerous authors [9], [12]-[14] on day-time, enlarging the time interval in which this occurs. The direct pharmacological effects of nicotine in different daily activities reduced during the night-time could explain this fact.

Furthermore, by linearly approximating the behaviour of the circadian rhythm in each of the four characteristic periods, it was possible to analyze as quickly the BP changed along 24h. In particular, in the first period the slopes were different among the four subject groups, showing different velocity changes of BP (both SBP and DBP) with higher values in both hypertensive and normotensive non-smokers than in smokers. On the contrary, in the successive three periods, the smokers had slopes generally greater than non-smokers in both hypertensive and normotensive subjects. The values were similar between smokers and non-smokers only during the third period for DBP in normotensive subjects and the fourth period, the awaking, for SBP in hypertensive subjects. A possible explanation of the latter behavior could be that during awakening, BP variations are mostly regulated by the autonomic nervous system that minimizes the effect of smoking.

The slopes could be used to quantify the morning blood pressure surge as well as the night blood pressure fall, associated with acute cardiovascular effects [21]. Hence, we can conclude that the different linear velocity rates of BP changes during 24h may be associated with different risk levels of cardiovascular disease. However, more clinical data are necessary to confirm this hypothesis.

## V. CONCLUSION

Our study allowed describing with an appropriate time resolution the circadian BP rhythm in smokers and non-smokers hypertensive and normotensive subjects. We highlighted significant differences between smokers and non-smokers in both hypertensive and normotensive subjects, especially during day-time. More interesting, the rates of change of both SBP and DBP in the four intervals were different in the four groups of subjects, higher in non-smokers between 9:00 and 14:30 and in smokers in the other three periods.

ACKNOWLEDGEMENT

The study was partially supported by Master in Clinical Engineering, University of Trieste.